\documentclass[twocolumn,pra,aps,showpacs,nofootinbib,tighten]{revtex4}
\usepackage{xspace,amsmath,amsfonts,amsthm,amssymb,amsbsy,graphicx,color}

\begin{document}


\preprint{LA-UR-01-1909}
\title{A simple formula for pooling knowledge about a quantum system}

\author{Kurt Jacobs}

\affiliation{Centre for Quantum Computer Technology, Centre for Quantum Dynamics, School of Science, Griffith University, Nathan 4111, Australia}

\begin{abstract}
When various observers obtain information in an independent fashion about a classical system, there is a simple rule which allows them to pool their knowledge, and this requires only the states-of-knowledge of the respective observers. Here we derive an equivalent quantum formula. While its realm of applicability is necessarily more limited, it does apply to a large class of measurements, and we show explicitly for a single qubit that it satisfies the intuitive notions of what it means to pool knowledge about a quantum system. This analysis also provides a physical interpretation for the trace of the product of two density matrices.  
\end{abstract}


\maketitle

An observer's information about a quantum system is captured by the density matrix that he or she assigns to it~\cite{note1}. The more certain the observer is about the state of the system, the more pure this density matrix; conversely, if he or she has no information regarding the quantum system, then the density matrix is maximally mixed, being proportional to the identity. 

Naturally, two observers need not have the same information about the quantum system, so that in general their density matrices for the system will be different. If two observers have obtained their information about a system independently, then together they have gathered more data about the system than they each have individually~\cite{note2}. The question then arises, is it possible for them to come up with a single density matrix, being a function only of their individual density matrices, which embodies their combined information?

For classical systems, in which the states-of-knowledge are probability densities, this is indeed the case when the observers have made their measurements without disturbing the system. In this case there is a simple formula for the combined state in terms of the individual states of knowledge. If the two states of knowledge concern the variable $x$, and are given by the probability densities $P_{\mbox{\scriptsize A}}(x)$ and $P_{\mbox{\scriptsize B}}(x)$, then the combined state is 
\begin{equation}
  P_{\mbox{\scriptsize AB}} = \frac{P_{\mbox{\scriptsize A}}(x) 
                                    P_{\mbox{\scriptsize B}}(x)}{\cal N} ,
  \label{cc}
\end{equation}
where ${\cal N}$ is the normalization. 

If the measurement data obtained about the system by the respective observers agrees, then their probability densities will be peaked about similar values of $x$, and the combined state of knowledge will have a lower entropy than the individual densities. This is to be expected, since the agreement of the two data sets serves to make the observer with both sets {\em more} certain of the value of $x$. However, if the two sets disagree significantly, then the combined density will have a higher entropy, since the observer will be less certain about the value of $x$. Further, the combined best-estimate for $x$ is more highly weighted towards that of the observer who was more certain to start with regarding the value of $x$ (that is, who's density was more sharply peaked), as one would expect. 

If the observers obtain their data in a consistent fashion (that is, they really do measure the same system), then it is more likely that their respective densities will agree, rather than disagree. In fact the  normalization, ${\cal N} = \int P_{\mbox{\scriptsize A}}(x)  P_{\mbox{\scriptsize B}}(x) dx$, is a measure of the relative probability that the two observers obtained their respective states of knowledge, and thus of how strongly they agree. It may therefore be viewed as a measure of the compatibility of the two states-of-knowledge. The concept of the compatibility of quantum states has been considered in references~\cite{BFM,CFS,BP}, and that of a {\em measure} for compatibility was introduced in~\cite{BP}. We will discuss this in more detail later. We note also that Poulin and Blume-Kohort have used this concept to propose a quantum formula for pooling knowledge, although the assumptions underlying this formula are quite different from those used to derive Eq.(\ref{cc}), and thus those which we will employ here~\cite{BP}. 

Our purpose here is to show that it is possible to derive a quantum formula for pooling knowledge by applying very similar conditions as those used to derive Eq.(\ref{cc}), although its range of applicability is more limited than the classical formula. To do this we first need to examine classical measurement theory and the assumptions which lead to Eq.(\ref{cc}). Classical measurements that cause no disturbance are described by the theory of Bayesian statistical inference~\cite{Bayes,Jaynes}. Since quantum measurement theory reduces to this theory when all states and operators are positive and commute (and thus may be taken as diagonal)~\cite{pool}, we can use the formalism of quantum measurement theory to treat such measurements~\cite{NC,Schumacher}. A classical disturbance-free measurement is thus a set of positive diagonal operators $\{E_k\}$ such that $\sum_k E_k = 1$. If the initial state of the system is the (diagonal) density matrix $\rho$, then on obtaining outcome $k$ the observer's state-of-knowledge becomes 
\begin{equation}
  \rho_k = \sqrt{E_k} \rho \sqrt{E_k}/(\mbox{Tr}[E_k\rho]) .
\end{equation}
Note that here we are assuming that the measurement is {\em efficient}, which means the observer has full information about the outcome $k$. For classical disturbance-free  measurements the distinction between efficient and inefficient measurements is not important, but in the quantum case it is. In general the observer can also perform reversible transformations on the system, being permutations of the states. If, on obtaining measurement result $k$, the observer performs permutation $T_k$, the final state of the system is
\begin{equation}
  \rho_k = \frac{ T_k \sqrt{E_k} \rho \sqrt{E_k} T_k^{\mbox{\scriptsize T}} }
                {\mbox{Tr}[E_k\rho]}.
\end{equation}
It is worth noting that including the $T_k$ allows one to describe classical measurements that cause disturbance as well as extract information~\cite{Hardy,WG}. 

We now consider the situation in which two observers (Alice and Bob) make measurements to obtain information about a system, and in which the following conditions are satisfied: 1. The observers obtain all their information in independent measurements, and 2. during this process neither of them disturbs the system. The first of these conditions means that the initial state of the system for both observers is proportional to the identity, since the observers share no information about the system prior to making measurements. The second condition removes the operators $T_k$. Let Alice make the measurement $\{E_k\}$ first. Immediately after this measurement her state of knowledge is $\rho_{\mbox{\scriptsize A}} = E_k/\mbox{Tr}[E_k]$ for some $k$. Bob then makes his measurement, $\{F_j\}$, at some later time. Since he has no knowledge of Alice's measurement result, his state-of-knowledge prior to his own measurement is given by averaging over Alice's possible results. This is $\sum_k P_k E_k/(\mbox{Tr}[E_k]) = \sum_k E_k /N = I/N$; that is, Alice's measurement leaves Bob's state-of-knowledge unchanged because his initial state is maximally mixed. However, had his state not been maximally mixed, Alice's measurement would still have left it unchanged, because her measurement is classical and therefore does not disturb the system. Bob's state-of-knowledge after his own measurement is then $\rho_{\mbox{\scriptsize B}} = F_j/\mbox{Tr}[F_j]$. 

Now we examine the state-of-knowledge of an observer with access to both Alice's and Bob's information (i.e. the results of both measurements). This is given by performing both measurements, one after the other, and the result is
\begin{eqnarray}
   \rho_{\mbox{\scriptsize AB}}   
            = \frac{ \sqrt{F_j} \sqrt{E_k} (I/N) \sqrt{E_k} \sqrt{F_j} }
                   { \mbox{Tr}[F_j E_k (I/N)] } 
            = \frac{ \rho_{\mbox{\scriptsize A}} \rho_{\mbox{\scriptsize B}} }
                   { \mbox{Tr}[\rho_{\mbox{\scriptsize A}} \rho_{\mbox{\scriptsize B}}]} .
  \label{c2}
\end{eqnarray}
The right hand side is the classical rule for pooling knowledge given in Eq.(\ref{cc}) above, but this time written using the quantum formalism: to pool their knowledge Alice and Bob simply multiply their probability densities together, and normalize the result. 
Since the measurements are classical all the measurement operators $\sqrt{E_k}$ and $\sqrt{F_j}$ commute, and so this formula holds just as well if Bob measures first. Also, because of the commutivity, Alice's state of knowledge, valid at time $t$, also applies just as well after Bob has made his measurement; its validity is unchanged until Alice makes another measurement. Further, this simple rule still holds if Alice and Bob have each made multiple measurements at a sequence of interspersed times. Because all the measurement operators commute, multiple interspersed measurements can always be separated out and described as a single measurement by each observer. 

To analyze the quantum case we first note that a quantum measurement is also described by a set of positive operators $E_k$, but the effect on the system is,  
\begin{equation}
  \rho_k = \frac{ A_k \rho A_k^\dagger } { \mbox{Tr}[E_k \rho] } ,
\end{equation}
for an efficient measurement, where $A_k^\dagger A_k = E_k$. If we use the polar decomposition theorem~\cite{NC}, then we can always write $A_k = U_k \sqrt{E_k}$, where $U_k$ is a unitary transformation, being the quantum equivalent of the reversible classical transformation $T_k$. The state of system after the measurement is
\begin{equation}
  \rho_k = \frac{ U_k \sqrt{E_k} \rho \sqrt{E_k} U_k^\dagger } 
                { \mbox{Tr}[E_k \rho] } .
\end{equation}
This has a very similar structure to the classical measurement -- a positive operator whose action changes the von Neumann entropy of the system, and thus describes the acquisition of information, followed by a reversible operation on the system. The similarity of the quantum and classical expressions suggests that one could view a quantum measurement in which $A_k = \sqrt{E_k}$ as the quantum equivalent of a classical disturbance-free measurement. Such measurements have been referred to in the literature as ``measurements without feedback''~\cite{FJ}, and it has been noted that there is some motivation for calling them ``minimally disturbing measurements''~\cite{Barnum}. Here we will use the more succinct term {\em bare measurements}. 

Now we impose the same conditions as in the classical analysis above. The initial density matrix is once again proportional to the identity, and the observers do not perform any unitary transformations on the system, so that the $U_k$ are set to the identity. We could view this as the condition that the observers refrain from disturbing the system any more than is strictly necessary to extract the information which they wish to obtain by their measurement. In addition to these we impose the further condition that Alice's and Bob's measurements are efficient. When alice makes the measurement $\{E_k\}$ at time $t$, and Bob the measurement $\{F_j\}$ later at $t+\tau$, then Alice's state of knowledge immediately after her measurement is (as above) $\rho_{\mbox{\scriptsize A}} = E_k/\mbox{Tr}[E_k]$ (for some $k$), and Bob's state of knowledge after his measurement is $\rho_{\mbox{\scriptsize B}} = F_j/\mbox{Tr}[F_j]$ for some $j$. The state of knowledge of an observer with access to the results of both measurements is 
\begin{eqnarray}
   \rho_{\mbox{\scriptsize AB}}   
            = \frac{ \sqrt{F_j} \sqrt{E_k} (I/N) \sqrt{E_k} \sqrt{F_j} }
                   { \mbox{Tr}[F_j E_k (I/N)] } 
            = \frac{ \sqrt{\rho_{\mbox{\scriptsize B}}} \rho_{\mbox{\scriptsize A}} 
                     \sqrt{\rho_{\mbox{\scriptsize B}}} }
                   { \mbox{Tr}[\rho_{\mbox{\scriptsize A}} \rho_{\mbox{\scriptsize B}}]}
  \label{qa}
\end{eqnarray}
We see from this that when Alice and Bob make efficient bare measurements there is a simple quantum rule that Alice and Bob can use to pool their knowledge. Note that the state that Alice uses in this rule, $\rho_{\mbox{\scriptsize A}}$, is the state she has immediately after her measurement, and does not take into account the fact that Bob has made a measurement. Two things should be noted in relation to this. The first is that neither Alice nor Bob need to know that the other has made a measurement. The second is that, because Alice's state does not take Bob's measurement into account, Alice's and Bob's states of knowledge will not, in general, be {\em consistent} states in the sense of~\cite{BFM}. This is because Alice's state refers to her knowledge at time $t$, and Bob's state to his knowledge at time $t+\tau$. 

Unlike the classical rule, the above quantum rule is not symmetric in the two states $\rho_{\mbox{\scriptsize A}}$ and $\rho_{\mbox{\scriptsize B}}$. In order to apply it, Alice and Bob must know the {\em time order} in which they made their measurements. However, we can easily use the above formula to obtain the quantum rule in the case when Alice and Bob know only their respective states of knowledge regarding a quantum system, and not the time ordering of their measurements. In this case they must average over both possibilities. The rule therefore becomes
\begin{eqnarray}
   \rho_{\mbox{\scriptsize AB}}   
            = \frac{ \sqrt{\rho_{\mbox{\scriptsize A}}} \, \rho_{\mbox{\scriptsize B}} \sqrt{\rho_{\mbox{\scriptsize A}}} + \sqrt{\rho_{\mbox{\scriptsize B}}} \rho_{\mbox{\scriptsize A}} \sqrt{\rho_{\mbox{\scriptsize B}}}}
                   { 2 \mbox{Tr}[\rho_{\mbox{\scriptsize A}}\rho_{\mbox{\scriptsize B}}]} .
  \label{qs}
\end{eqnarray}
We consider this, rather than Eq.(\ref{qa}), as the quantum equivalent of the classical rule for pooling knowledge because this symmetric formula, like its classical counterpart, refers only to the states of knowledge possessed by each observer, and not to any other information, such as the time ordering of their measurements. We now discuss various properties of this formula:

1. Both the classical and quantum formulae are only applicable for certain classes of measurements. In the classical case these are those measurements which do not disturb the system at all. Since in the classical world it is common to be able to make disturbance-free measurements, it is natural to consider this class of measurements, and to assume that the result is widely applicable. In the quantum case, the restriction is to efficient bare measurements, and since there is not as clear a separation between these measurements and other kinds of quantum measurements, one might expect the quantum formula to be much less widely applicable. However, measurements that describe purely the extraction of information about an observable (a Hermitian operator) are bare, and the formula does apply to this large and important class, as well as many others.

2. The classical formula is also more widely applicable than the quantum formula because it applies to situations in which any number of measurements have been made by the two observers, and these measurements can be interspersed in time with one another. When one combines two bare measurements, by performing one after the other, the result is, in general, a measurement which is not bare. Thus, the quantum result does not apply when Alice's and Bob's measurements are interspersed. It does apply when Alice makes a sequence of efficient measurements, the overall result of which is bare, and then some time later Bob makes a sequence of such measurements, the overall result of which is also a bare measurement. This would be the case, for example, if Alice were to make a sequence of efficient measurements all of which extract information about the same observable, and then Bob made a second sequence, all of which extract information about a second observable.

3. The quantum formula reduces to the classical formula when all measurements commute.

4. The formula involves the expression $\mbox{Tr}[\rho_{\mbox{\scriptsize A}}\rho_{\mbox{\scriptsize B}}]$. This is not the fidelity between $\rho_{\mbox{\scriptsize A}}$ and $\rho_{\mbox{\scriptsize B}}$, and does not satisfy the properties required of a fidelity~\cite{Josza}. 
However, the above analysis provides a physical interpretation for $\mbox{Tr}[\rho_{\mbox{\scriptsize A}}\rho_{\mbox{\scriptsize B}}]$: it is the ``normalized'' probability that Bob will obtain the state-of-knowledge $\rho_{\mbox{\scriptsize B}}$ given that Alice has the state $\rho_{\mbox{\scriptsize A}}$. What we mean by this is the following: When Bob makes an efficient measurement on a system which Alice knows to be in state $\rho_{\mbox{\scriptsize A}}$, and obtains the result associated with the operator $E_k$ (recall that in the knowledge pooling scenario Bob's state is $\rho_{\mbox{\scriptsize B}} = E_k/\mbox{Tr}[E_k]$), then the probability that Bob obtains this result is $P = \mbox{Tr}[E_k\rho_{\mbox{\scriptsize A}}]$. However, for a given $\rho_{\mbox{\scriptsize B}}$ we can chose this probability to be as small as we like, by scaling $E_k$ so that its norm is arbitrarily small. Thus, to compare the relative probability for Bob to have various different states-of-knowledge given Alice's state-of-knowledge, we should chose a consistent value for the norm of $E_k$. The expression $\mbox{Tr}[\rho_{\mbox{\scriptsize A}}\rho_{\mbox{\scriptsize B}}] = P/\mbox{Tr}[E_k]$ is such a normalized version of $P$. Hence $\mbox{Tr}[\rho_{\mbox{\scriptsize A}}\rho_{\mbox{\scriptsize B}}]$ can be viewed as a measure of the compatibility of the two states-of-knowledge. That is, if $\mbox{Tr}[\rho_{\mbox{\scriptsize A}}\rho_{\mbox{\scriptsize B}}]$ is small, then it is unlikely that these two states would arise together, whereas if it is large, it is likely that two observers, making measurements but interfering with the system as little as possible, would hold these states. However, we should be clear about the context of this compatibility -- as noted earlier, since Alice's and Bob's states do not take into account that the other may have made a measurement at a later time, their states are not consistent states under the definition of compatibility used in~\cite{BFM} and~\cite{BP}. As a result $\mbox{Tr}[\rho_{\mbox{\scriptsize A}}\rho_{\mbox{\scriptsize B}}]$ does not measure the notion of compatibility considered there. 

The preceding analysis also applies to the classical case, and explains our previous assertion that the normalization, ${\cal N}$, in Eq.(\ref{cc}) provides a measure of the compatibility of the two probability densities. Note also that in the classical case the observers' states are consistent in the sense of~\cite{BFM} and~\cite{BP}, and thus the measure ${\cal N}$ does apply to the notion of compatibility considered there. 

5. Let us examine the behavior of the formula when the system in question is a single qubit. In this case $\rho_{\mbox{\scriptsize A}}$ and $\rho_{\mbox{\scriptsize B}}$ can be represented by the Bloch vectors ${\bf a}$ and ${\bf b}$, so that $\rho_{\mbox{\scriptsize A}} = 1/2 (I + {\bf a}\cdot{\boldsymbol \sigma})$ and $\rho_{\mbox{\scriptsize B}} = 1/2 (I + {\bf b}\cdot{\boldsymbol \sigma})$ where ${\boldsymbol \sigma} = (\sigma_x,\sigma_y,\sigma_z)$ is the vector of Pauli matrices. Using the multiplication rule for the Pauli matrices, $({\bf a}\cdot{\boldsymbol \sigma}) ({\bf b}\cdot{\boldsymbol \sigma})  = ({\bf a}\cdot{\bf b}) I + i{\boldsymbol \sigma} ({\bf a}\times {\bf b}) $, one can calculate the combined state of knowledge. In this case, $\mbox{Tr}[\rho_{\mbox{\scriptsize A}}\rho_{\mbox{\scriptsize B}}] = (1/2)(1 + {\bf a}\cdot{\bf b})$, and the combined state of knowledge is  
\begin{equation}
 \rho_{\mbox{\scriptsize AB}} = \frac{1}{2} ( I + \frac{1}{\mbox{Tr}[\rho_{\mbox{\scriptsize A}}\rho_{\mbox{\scriptsize B}}]}(\alpha {\bf a} + \beta {\bf b}) \cdot {\boldsymbol \sigma}) 
\label{qubit}
\end{equation}
where 
\begin{eqnarray}
  \alpha & = & \frac{1}{2} + \frac{1}{4}\left( \frac{{\bf a}\cdot{\bf b}}{f(a)} - \frac{b^2}{f(b)}\right) , \\
  \beta & =  & \frac{1}{2} + \frac{1}{4}\left( \frac{{\bf a}\cdot{\bf b}}{f(b)} - \frac{a^2}{f(a)}\right)  ,
\end{eqnarray}
and where $a \equiv |{\bf a}|$, $b \equiv |{\bf b}|$ and $f(x) \equiv 1 + \sqrt{1 - x^2}$. The bloch vector for the combined state of knowledge is therefore a weighted sum of the bloch vectors of the individual states. We note that this is not necessarily a weighted {\em average} of ${\bf a}$ and ${\bf b}$ because $\alpha + \beta$ will in general not be equal to $\mbox{Tr}[\rho_{\mbox{\scriptsize A}}\rho_{\mbox{\scriptsize B}}]$. Now what properties do we expect this formula to have? First we note that the length of the Bloch vector is a measure of how certain each observer is about the state of the system. Therefore, when $a=b$, there is no reason to believe Alice's claim about the system any more than Bob's, and hence the combined state should be simply the average of the two Bloch vectors. It is clear by inspection that this is the case. Second, we would expect that when the two initial Bloch vectors point in the same direction, the resulting state should have a longer Bloch vector than either of the two initial vectors. This is because the observer with the combined knowledge should be more certain of the state of the system than either Alice or Bob. This is equivalent to the statement that ${\bf a}\cdot {\bf b} = ab$ implies that $\alpha a + \beta b \geq \mbox{Tr}[\rho_{\mbox{\scriptsize A}}\rho_{\mbox{\scriptsize B}}] \mbox{max}(a,b)$, and it is straightforward to show that this is true. Finally, we would expect that the direction of the Bloch vector for the combined state should be closer to that of the initial Bloch vector which is the longer of the two, since the final state should agree more with the state of the observer who has the higher degree of certainty. This is equivalent to the statement that $a \geq b$ implies that $\alpha a \geq \beta b$. This is also quite straightforward to show, and follows from the fact that $g(x) = x/(1 + \sqrt{1 - x^2})$ is monotonically increasing for $x \in [0,1]$. Thus we see explicitly for the case of a single qubit that the formula satisfies the intuitive notions of what it means to pool knowledge regarding a quantum system. 

5. The formula generalizes in a straightforward manner to $N$ observers, each making an efficient bare measurement at a set of mutually distinct times. For example, for three observers, none of whom have information regarding the times at which their information was obtained, the formula is
\begin{equation}
\rho_{\mbox{\scriptsize ABC}} = \frac{ \sum_{\mbox{\scriptsize }} 
                                \sqrt{\rho_{\mbox{\scriptsize A}}}
                                      \sqrt{\rho_{\mbox{\scriptsize B}}}
                                            \rho_{\mbox{\scriptsize C}}
                                      \sqrt{\rho_{\mbox{\scriptsize B}}}  
                                      \sqrt{\rho_{\mbox{\scriptsize A}}} 
                                     }{
                            3! \, \mbox{Tr}[\rho_{\mbox{\scriptsize A}} 
                                            \rho_{\mbox{\scriptsize B}}
                                            \rho_{\mbox{\scriptsize C}}] 
                                      }
\end{equation}
where the sum is over all permutations of A, B and C.

In conclusion, it is possible to obtain a formula for pooling knowledge about classical systems  which requires only the states of knowledge of the respective observers. This is possible when the observers make disturbance-free measurements. We have shown that it is also possible to derive a quantum formula for pooling knowledge, which, although its applicability is more limited, also requires only the states-of-knowledge of the respective observers. In the quantum case this is possible when the observers make efficient bare measurements. In this setting it is therefore these measurements which are the natural equivalent of classical disturbance-free measurements. Since bare measurements include those whose sole action is to extract information about a single observable (which includes continuous measurements of an observable integrated over finite times), the formula applies to a wide class of measurements. 

%

{\em Acknowledgements:} The author would like to thank Robin Blume-Kohort, David Poulin, Rob Spekkens, Howard Wiseman and Bill Wootters for helpful and stimulating discussions regarding knowledge pooling and quantum state compatibility, and Howard Wiseman for a critical reading of the manuscript.


\begin{thebibliography}{99}
\bibitem{note1} 
Specifically, the density matrix characterizes the observers knowledge regarding the likely results of all future measurements which can be made on the system. It is this knowledge with which we will be concerned here.
\end{thebibliography}
\end{document}